\documentstyle{article}
\title{On  Asymptotic Hamiltonian for SU(N) Matrix Theory    }
\author{Anatoly Konechny \\ Department of Mathematics, University of California, \\ 
Davis, CA 95616 \\ konechny@math.ucdavis.edu } 
\date { October 16, 1998} 
\begin{document}
\maketitle 
\smallskip
\begin{abstract} 
We compute the leading contribution to the effective Hamiltonian of 
 $SU(N)$ matrix theory  in the limit of large separation. We work with a gauge fixed Hamiltonian 
and use  generalized Born-Oppenheimer approximation, 
extending the recent work of Halpern and Schwartz for $SU(2)$ \cite{HS}.
The answer turns out to be a free 
Hamiltonian for the coordinates along the flat directions of the potential. Applications to finding  
 ground state candidates and calculation of the correction (surface) term to Witten  index 
are discussed. 

\end{abstract}
\section{Introduction}
We study the maximal supersymmetric gauge quantum mechanics. 
This model was first considered in papers \cite{CH}, \cite{Flume}, \cite{BRR}.  
 Later  this Hamiltonian found applications in the physics of  supermembranes 
\cite{deWHN}, \cite{deWLN}  and $D$-particles \cite{Witten}. 
There is a remarkable conjecture, called Matrix theory, that the same Hamiltonian gives 
a nonperturbative description of M-theory \cite{BFSS}. This conjecture implies the 
existence of a unique bound state at threshold for  each $SU(N)$ gauge group where $N\ge 2$ 
(though the existence of the state was conjectured earlier). 

The $SU(2)$ case was recently studied in a number of papers  
\cite{HS}, \cite{SS}, \cite{Yi}, \cite{PfW}. 
In \cite{SS} it was proved, using the computation of Witten  index, that at least one ground state 
exists in $SU(2)$ model.  Some progress also have been achieved in the general $SU(N)$ case 
\cite{PR}, \cite{MNS}, \cite{Hoppe}. 
In \cite{MNS} the principal contribution to Witten index was computed (see also \cite{GG1} and 
 \cite{GG2}). Calculation of the surface 
contribution to the index was performed in \cite{GG1} under certain assumptions. The present paper 
justifies these assumptions.
 In \cite{HS} an asymptotic form of a  wave function was proposed as a candidate ground 
state.

The present paper extends the ideas of \cite{HS} and \cite{PfW} to  higher $N$'s. 
 We consider the model in the asymptotic region where the coordinates along 
the flat directions of the potential become large. Our method is generalized Born-Oppenheimer 
approximation introduced in \cite{HS}. The authors of \cite{HS} tackle the problem of 
gauge invariance by working  with a complete set of gauge invariant variables. It would be rather hard 
to implement this idea for higher $N$'s. Instead, we apply the gauge fixing procedure of \cite{deWLN}. 
The result of our computation is that in the leading order the dynamics of flat coordinates is 
described 
by the free Hamiltonian. Possible potential term  vanishes due to  cancellations between 
many different terms. This result is not surprising, it is generally expected in supersymmetric theories. 
It is in  agreement with $SU(2)$ calculations of    \cite{HS}, \cite{PfW}, 
\cite{SS}. In the last section of the paper we discuss possible applications of the main result.

\section{The model and preliminaries} 
In this section we introduce the model and do some preliminary technical work needed for the Born-Oppenheimer 
approximation. Namely, we write down the (partially) gauge fixed Hamiltonian, 
split it into 4 basic parts: free Hamiltonian for slow (Cartan) degrees of freedom,  bosonic and 
fermionic oscillators, and the interaction part. 
After that we explain the quantization of the oscillators part. 
We refer the reader to the seminal 
paper \cite{deWLN} for all the details of the  gauge fixing procedure.

The Hamiltonian of  matrix theory is that of the 10D super Yang-Mills 
theory dimensionally reduced to 0+1 dimensions (in $A_{0}=0$ gauge). 
 One can write it in the following form
\begin{equation} \label{H}
H=\frac{1}{2}\pi^{\mu}_{A} \pi^{\mu}_{A} + 
\frac{1}{4}f_{ABC}\phi^{\mu}_{B}\phi^{\nu}_{C}f_{ADE}\phi^{\mu}_{D}\phi^{\nu}_{E} 
-\frac{i}{2}f_{ABC}\phi^{\mu}_{A}\Lambda_{B\alpha}\Gamma^{\mu}_{\alpha \beta}\Lambda_{C\beta} 
\end{equation}
where $\phi^{\mu}_{A}$, $\pi^{\mu}_{A}$ are   real bosonic variables and $\Lambda_{A \alpha}$ are  real 
fermionic variables. 
The lower capital Latin indices are that of the adjoint representation of a real compact Lie algebra $\bf g$ with 
  totally antisymmetric structure constants $f_{ABC}$. Denote by $G$ the Lie group of $\bf g$.
 The indices $\mu, \nu =1, \dots , 9$, $\alpha=1, \dots , 16$ correspond to vector and real spinor representations 
of the group $spin(9)$ respectively. 
The $SO(9)$ gamma matrices $\Gamma^{\mu}$ are assumed to be real and symmetric. 
The canonical commutation relations between the variables are 
$$ 
[\pi^{\mu}_{A}, \phi^{\nu}_{B}]=-i\delta^{\mu \nu}\delta_{AB} \, , \quad
\{\Lambda_{A\alpha}, \Lambda_{B\beta}\} = \delta_{\alpha \beta}\delta_{AB} .
$$
From now on we set $G=SU(N)$ that corresponds to the Matrix theory with the ``center of mass'' 
degrees of freedom being excluded. The Lie algebra ${\bf g}=su(N)$ consists of the traceless antihermitean 
$N\times N$ matrices $\phi$. An invariant, positive definite inner product on $\bf g$ is defined as 
$(\phi_{1}, \phi_{2})=-2Tr(\phi_{1}\phi_{2})$. The Cartan subalgebra consists then of  
diagonal matrices $\phi \in {\bf g}$. Let  matrices $T_{A} , \enspace A=1, \dots , N^{2}-1$
 be an orthonormal 
basis in $su(N)$ with respect to this inner product such that the indices $A=1, \dots , N-1$ 
correspond to the Cartan subalgebra. Then 
 $f_{ABC}=(T_{A}, [T_{B}, T_{C}])$.

Now we will briefly explain the gauge fixing procedure. 
 Given any element $\phi \in g$ there exist a unique 
matrix $D$ such that $\phi=UDU^{-1}$ for some $U\in G$ and $D$ is of the form 
\begin{equation} \label{D}
D=i\left( \begin{array}{cccc} 
\lambda_{1} & 0 & \ldots & 0 \\
0&\lambda_{2}  & \ldots &0 \\
\vdots & \vdots & \ddots & \vdots \\
0&0& \ldots &\lambda_{N} 
\end{array}
\right)
\end{equation}
where $\lambda_{n}$ are real numbers such that 
\begin{equation}
\sum_{n=1}^{N}\lambda_{n} = 0 \, , \quad \lambda_{1} \ge \lambda_{2} \ge \ldots \ge \lambda_{N} . 
\end{equation}
The transformation $U$ is defined up to multiplications by an arbitrary element of the Cartan subgroup. 
Thus, it is clear that we can perform a partial gauge fixing by requiring that $\phi^{9}$ 
lies within the Cartan subalgebra and has the form (\ref{D}) of  the matrix $D$ above. This particular form corresponds 
to a fixed  Weyl chamber within the Cartan subalgebra.   
The residual gauge group  is then $U(1)^{N-1}$.  
This gauge fixing procedure is described in details in \cite{deWLN}. 
Following \cite{deWLN} we adopt the convention that 
the indices $i,j,k, \dots$ are summed from $1$ to $N-1$ (the indices of Cartan subalgebra) 
, while capital indices    $I,J,K, \dots $ from the middle of the alphabet run from $N$ to $ N^{2}-1$ 
(the indices of the subspace spanned by the roots). Also we assume that the indices $a,b,c, \dots$ 
run from $1$ to $8$ and correspond to the first eight coordinates in $R^{9}$. Finally,  let us  adopt a new 
notation for the bosonic Cartan variables $D_{i}^{\mu}\equiv \phi_{i}^{\mu}$ 
(that will be convenient in the future when we split the variables into the ``fast'' and ``slow'' ones). 
The gauge fixed Hamiltonian is $H=H_{0} + H_{B} + H_{F} + H_{4}$, where 
\begin{equation} \label{H0}
H_{0}=-\frac{1}{2}\frac{\partial^{2}}{\partial D_{i}^{\mu}\partial D_{i}^{\mu}} 
\end{equation}
\begin{equation} \label{HB}
H_{B}=-\frac{1}{2}\frac{\partial^{2}}{\partial \phi_{I}^{a}\partial \phi_{I}^{a}} + 
\frac{1}{2}\Omega^{2}_{IJ}\phi_{I}^{a}\phi_{J}^{a} 
\end{equation}
\begin{equation} \label{HF}
H_{F}=-\frac{i}{2}\Lambda_{I}z^{\mu}_{IJ}\Gamma^{\mu}\Lambda_{J}
\end{equation}
\begin{eqnarray} \label{H4}
\lefteqn{H_{4}=\frac{1}{4}f_{AIJ}f_{AKL}\phi_{I}^{a}\phi_{J}^{b}\phi_{K}^{a}\phi_{L}^{b} +  
 \frac{1}{2}f_{AiJ}f_{AKL}(D^{a}_{i}\phi_{J}^{b}-D_{i}^{b}\phi_{J}^{a}) \phi_{K}^{a}\phi_{L}^{b} - } \nonumber \\
&& -\frac{1}{2}f_{AiJ}f_{AkL}D_{i}^{a}D_{k}^{b}\phi_{J}^{b}\phi_{L}^{a} - 
\frac{i}{2}f_{IAB}\phi_{I}^{a}\Lambda_{A}\Gamma^{a}\Lambda_{B} + 
\frac{1}{2}(w^{t}w)_{IJ}\hat L_{I}\hat L_{J} 
\end{eqnarray}
The following notations are used above: 
$$
 \Omega^{2}_{IJ} = (z_{\mu}^{t}z_{\mu})_{IJ} \, , \quad z_{IJ}^{\mu}=D_{i}^{\mu}f_{iIJ} \,  , \quad w= (z^{9})^{-1}
$$
$$
\hat L_{I}=-f_{IBC}\left( i\phi_{B}^{a}\frac{\partial}{\partial \phi_{C}^{a}} + 
\frac{i}{2}\Lambda_{B\alpha}\Lambda_{C\alpha} \right) .
$$  
This Hamiltonian is self-adjoint with respect to the measure 
$\prod_{a,A}d\phi^{a}_{A}\prod_{i}dD_{i}^{9}$ (the Faddeev-Popov Determinant was eliminated by redefining 
the Hilbert space).

The terms $H_{B}$ and $H_{F}$ are Hamiltonians of bosonic and fermionic harmonic oscillators. 
To diagonalize these Hamiltonians,  note that the eigenvectors of the matrices $z_{IJ}^{\mu}$ are the 
complex root vectors $E_{mn}^{I}$ ($m,n=1 , \ldots , N \, , m\ne n$). These eigenvectors satisfy  
\begin{equation} \label{roots1} 
z_{IJ}^{\mu}E_{mn}^{J}=i(\lambda_{m}^{\mu}-\lambda_{n}^{\mu})E_{mn}^{I} 
\end{equation}
\begin{equation}
(E_{mn}^{I})^{*}=E_{nm}^{I} .
\end{equation}
Here $\lambda_{m}^{\mu}$ are eigenvalues of  matrices $D^{\mu}=\sum_{i}D_{i}^{\mu}T^{i}$. 
The vectors $E_{mn}$ define an orthonormal basis of a root subspace: 
\begin{equation}
\sum_{I} (E_{mn}^{I})^{*}E_{pq}^{I}=\delta_{mp}\delta_{nq} \quad (m\ne n , p\ne q ) 
\end{equation}
\begin{equation}
\sum_{m\ne n}(E_{mn}^{I})^{*}E_{mn}^{J} = \delta_{IJ} .
\end{equation}
Note the equality 
$E_{mn}\equiv E_{mn}^{I}T^{I}=\frac{i}{\sqrt{2}}e_{mn}$ , where $e_{mn}$ is $N\times N$ matrix 
with $1$ on the $m,n$-th place and zeros elsewhere. These matrices satisfy 
the following commutation relations 
\begin{equation} \label{roots5}
[E_{mn}, E_{pq}] = \frac{i}{\sqrt{2}}(\delta_{np}E_{mq} - \delta_{mq}E_{pn}) .  
\end{equation} 
The eigenvalues of $H_{B}$ and $H_{F}$ (as we will see further) are 
\begin{equation}
r_{mn}=\sqrt{\sum_{\mu=1}^{9}(\lambda_{m}^{\mu}-\lambda_{n}^{\mu})^{2}} . 
\end{equation}

Thus,  we can introduce new variables $\phi_{mn}^{a}$, $\Lambda^{mn}_{\alpha}$, 
$ m,n =1, \dots ,N , \, m\ne n  $ so that 
\begin{equation} 
\phi_{I}^{a}=\sum_{m\ne n} \phi_{mn}^{a}E_{mn}^{I} \, , \quad 
\Lambda_{I\alpha}=\sum_{m\ne n} \Lambda^{mn}_{\alpha}E_{mn}^{I} . 
\end{equation}
The commutation relations for fermions are now 
$$
\{\Lambda^{mn}_{\alpha}, \Lambda^{pq}_{\beta}\}=\delta_{\alpha\beta}\delta^{np}\delta^{mq}
$$
and it is natural to set $(\Lambda^{mn})^{\dagger}=\Lambda^{nm} \, m<n$. 
However,  to diagonalize $H_{F}$ some more work needs to be done. 
Namely, one should use the $spin(9)$-rotated fermions 
$$
\tilde \Lambda^{mn}_{\alpha} = R^{mn}_{\alpha\beta}\Lambda^{mn}_{\beta} \, m<n 
$$
$$
\tilde \Lambda^{mn}_{\alpha} = R^{nm}_{\alpha\beta}\Lambda^{mn}_{\beta} \, m>n 
$$
where
$$
R^{mn}_{\alpha\beta}=
\frac{r_{mn} + (\lambda_{m}^{\mu} - \lambda_{n}^{\mu})\Gamma^{9}\Gamma^{\mu}}
{ 2r_{mn}(r_{mn} + \lambda_{m}-\lambda_{n})} \enspace m<n
$$
is an orthogonal matrix. 
Here and everywhere $\lambda_{n}$ with the suppressed upper index stands for $\lambda_{n}^{9}$.
In terms of $\tilde \Lambda^{mn}$ variables $H_{F}$ can be written as 
\begin{equation} \label{HF2}
H_{F}=\sum_{m<n}r_{mn}\left( (\tilde \Lambda_{+}^{mn})^{\dagger}\tilde \Lambda_{+}^{mn} + 
\tilde \Lambda_{-}^{mn}(\tilde \Lambda_{-}^{mn})^{\dagger} - 8 \right)
\end{equation}
where $\Lambda_{\pm} $ denotes the chiral components taken with respect to $\Gamma^{9}$.

Using (\ref{roots1}) - (\ref{roots5}) we can rewrite all parts of the Hamiltonian  
in terms of the variables 
$D_{i}^{\mu}$, $\Lambda_{i\alpha}$, $\phi^{a}_{mn}$, $\Lambda_{\alpha}^{mn}$. 
The corresponding expressions for $H_{0}$ and $H_{F}$ are given in (\ref{H0}) and (\ref{HF2}). 
The Hamiltonian for bosonic oscillators reads now as 
\begin{equation} \label{HB2}
H_{B}=-\sum_{m<n}\frac{\partial^{2}}{\partial\phi_{mn}^{a}\partial (\phi_{mn}^{a})^{*}} + 
\sum_{m<n}r_{mn}^{2}\phi^{a}_{mn}(\phi_{mn}^{a})^{*}  .
\end{equation}
As one can easily see from (\ref{HF2}) and (\ref{HB2}) the ground state energies for the bosonic and 
fermionic oscillators precisely cancel each other (we have $8$ bosonic modes for each $r_{mn}, m < n$). 
The expression for $H_{4}$ in terms of the new variables is rather long.  We relegate it  to the appendix A
in order not to  complicate the main text.

The normalized state vector of the oscillators 
ground state has the following form  
\begin{eqnarray} \label{ground} 
\lefteqn{ |0\rangle =|0_{B}\rangle |0_{F}\rangle } \nonumber \\
  |0_{B}\rangle &=&
\left( \prod_{m<n}r_{mn}^{4}2^{4}\pi^{-4} \right) 
exp\left( -\sum_{m<n}r_{mn}\phi_{mn}^{a}\phi_{nm}^{a} \right)
\nonumber \\
|0_{F}\rangle &=& \prod_{m<n} \prod_{\alpha=1}^{8}\left( \tilde \Lambda_{-\alpha}^{mn} \right)^{\dagger}|Fock\rangle
\end{eqnarray}
where $|Fock\rangle$ denotes the Fock vacuum for the fermions $\Lambda^{mn}$. 
It is not hard to check that (\ref{ground}) is invariant under the residual $U(1)^{N-1}$ gauge transformations. 


\section{Born-Oppenheimer approximation and perturbation theory} 
The important feature of Matrix model that makes the existence of a threshold bound state 
possible is flat directions of the potential. Under the gauge-fixing condition that we employed, the 
coordinates along the flat directions are Cartan variables $D_{i}^{\mu}$. We group them together with 
their fermionic counterparts $\Lambda_{i}$. The coordinates in transverse directions are $\phi^{a}_{mn}$ 
along with their superpartners $\Lambda^{mn}$. Following the terminology of 
Born-Oppenheimer approximation we call the variables $D_{i}^{\mu}$, $\Lambda_{i}$ ``slow'' and 
the variables $\phi^{a}_{mn}$ , $\Lambda^{mn}$ ``fast''. When slow variables are considered to be
fixed the dynamics of the fast ones is governed by the oscillator Hamiltonians $H_{B}+H_{F}$ and 
by $H_{4}$.
In the asymptotic region where the frequencies $r_{mn}$ of the oscillators are large and assuming 
that $H_{4}$ can be treated as a perturbation, it is natural to expect that  the fast degrees 
of freedom will remain in the oscillators ground state. 
The right assymptotic region for our purposes turns out to be $D_{i}^{9}\to \infty$ . 
This is the same as setting $\lambda_{m}-\lambda_{n} \to \infty$ for all $m<n$ . 
We will assume that $\lambda_{m}- \lambda_{n} \, , m<n $ are all of the order $r$ where $r \to \infty$. 
Taking into account that the variables $\phi_{mn}^{a}$ in the oscillators ground state are 
of the order $1/\sqrt{r_{mn}}$ one can estimate that $H_{4}$ is of the order $1/\sqrt{r}$, that indeed 
allows one to treat it as a perturbation when $r\to \infty$.

Thus, we search for an approximate solution to the spectral  problem 
$$ (H-E)|\Psi\rangle=0 $$ in the form $ |\Psi\rangle=| \cdot\rangle|\Psi(D_{i}, \Lambda_{i})\rangle $ where 
$ |\cdot\rangle$ is the ground state 
of superoscillators. The general formalism for this problem was developed in \cite{HS}. We refer the reader for 
the detailed explanation of such generalized Born-Oppenheimer approximation to that paper. Here we 
just want to explain the main idea of the formalism and develop a perturbative expansion suitable for  the problem
at hand. If one introduces a pair of projection operators $P=|\cdot\rangle \langle \cdot|$, $Q=1-P$, 
the Schrodinger equation breaks into a system of two equations 
\begin{eqnarray}\label{projeq}  P(H-E)P|\Psi\rangle + P(H-E)Q|\Psi\rangle &=& 0 \nonumber \\ 
 Q(H-E)P|\Psi\rangle + Q(H-E)Q|\Psi\rangle &=& 0 . 
\end{eqnarray}  
The second equation can be formally solved as 
\begin{equation} \label{QPsi} 
Q|\Psi\rangle=-(Q(H-E)Q)^{-1}Q(H-E)P|\Psi \rangle . 
\end{equation}  
Substituting this solution into the first equation,  we get 
\begin{equation} \label{redS} 
 [P(H-E)P - P(H-E)Q(Q(H-E)Q)^{-1}Q(H-E)]P|\Psi \rangle \, ,
\end{equation}
i.e. the ``reduced'' Schrodinger equation for $P|\Psi\rangle=|\cdot\rangle|\Psi(D_{i}, \Lambda_{i})\rangle$. 
The first term in (\ref{redS}) is the conventional effective Hamiltonian for the Born-Oppenheimer approximation. 
 The second one constitutes the correction term. 
Now note that in the problem at hand $|\cdot\rangle$ is the ground state of superoscillators that has a zero energy. 
If we split the total Hamiltonian as $H=H_{osc} + H'$ where $H_{osc}$ is the superoscillators Hamiltonian  
and $H'$ is all the rest, then $H_{osc}|\cdot\rangle=0$.  A direct analysis of 
formulae (\ref{H0})-(\ref{H4}) shows that $QH_{osc}Q=Q(H_{B}+H_{F})Q$ scales like $r$ and $QH'Q$ 
scales like $O(1)$
(because of the part of $H_{0}$ that depends on $D^{a}_{i}$ variables). 
This is the reason to treat $QH'Q$ as a perturbation. 
 Now we can apply a perturbation theory in $QH'Q$ to the term $(Q(H-E)Q)^{-1}$ in 
(\ref{redS}) : 
\begin{eqnarray} \label{pert1}
\lefteqn{(Q(H-E)Q)^{-1} = (QH_{osc}Q + QH'Q - E)^{-1} = } \nonumber \\
&=& \frac{1}{QH_{osc}Q-E} -  
\frac{1}{QH_{osc}Q-E}QH'Q\frac{1}{QH_{osc}Q-E} + \ldots 
\end{eqnarray} 
Here $Q(H-E)Q$ is understood as an operator on the $Q$-projected subspace of the whole Hilbert 
space and thus operator inverse makes sense (by the same reason we can write $E$ instead of 
$EQ$ in this equation). 
 Substituting this perturbation expansion into  reduced Schrodinger equation (\ref{redS}), we get 
the following expression for the effective Hamiltonian
\begin{eqnarray} \label{perturb}
&&\langle\cdot|H'|\cdot\rangle - \langle\cdot|H'Q\frac{1}{QH_{osc}Q-E}QH'|\cdot\rangle + \nonumber \\
&&+ \langle\cdot|H'Q\frac{1}{QH_{osc}Q-E}QH'Q\frac{1}{QH_{osc}Q-E}QH'|\cdot\rangle + 
\ldots 
\end{eqnarray} 
The ``propagator'' $\frac{1}{QH_{osc}Q-E}$ scales like $1/r$ and the last expansion  
can be used to compute the effective Hamiltonian to any desired order in $1/r$. 
Although formally $H'$ scales like $O(1)$ because of the $H_{0}$ term, 
the contribution of the order $1/r^{2}$  comes solely from 
the first two terms in (\ref{perturb}). This happens due to the fact that $H_{0}|\cdot\rangle$ scales like $1/r$  
(see  appendix B). 
Hence, the correction term we need to have the effective Hamiltonian up to  the order $1/r^{2}$ is 
\begin{equation} \label{corr} 
- \langle\cdot|H_{4}Q\frac{1}{QH_{osc}Q-E}QH_{4}|\cdot\rangle  .
\end{equation}

Another way to get the correction term is  using an ansatz method similar to the one 
developed in \cite{HS}.  Both (\ref{corr}) and the ansatz method give the same result. 


\section{Calculation} 
 
First we need to compute the main contribution $\langle\cdot|H|\cdot\rangle = 
\langle\cdot|H_{0}|\cdot\rangle + \langle\cdot|H_{4}|\cdot\rangle$
up to the second order in $1/r$.  For  calculations  
it is convenient to express $H_{0}$ in terms of $\lambda_{n}^{\mu}$ variables. We have a linear 
correspondence between the variables $D^{\mu}_{i}=A_{i}^{n}\lambda_{n}^{\mu} $. Since 
by definition $\sum_{i}D_{i}^{\mu}T^{i} = i\sum_{i} \lambda_{i}e_{ii}$ the condition
 $(D^{\mu}, D^{\nu}) =-2tr(D^{\mu}D^{\nu})=D^{\mu}_{i}D^{\mu}_{i}$ gives 
$\lambda_{n}^{\mu}=\frac{1}{2}A^{i}_{n}D_{i}^{\mu}$. Therefore 
$$
H_{0} = -\frac{1}{4}\frac{\partial^{2}}{\partial \lambda_{n}^{\mu}\partial \lambda_{n}^{\mu}} 
$$
where the last operator is considered as a restriction to the subspace of functions  of 
$\lambda_{1}^{\mu} , \ldots , \lambda^{\mu}_{N}$ which are annihilated  
by $\sum_{n}^{N} \frac{\partial}{\partial \lambda_{n}^{\mu}} $ 
(which means that they depend only on differences 
$\lambda_{m}^{\mu}-\lambda_{n}^{\mu}$ ). Both bosonic and fermionic  oscillators ground state vectors 
 depend on $\lambda_{n}^{\mu}$ variables. This should be taken into account when  calculating  averages of 
differential operators such as $H_{0}$. Some useful formulae for computing such averages are given in 
 appendix B.  
After this preliminary work the direct computation yields
\begin{equation} \label{H0c}
\langle\cdot|H_{0}|\cdot\rangle = -\frac{1}{2}\frac{\partial^{2}}{\partial D_{i}^{\mu}\partial D_{i}^{\mu} } + 
 9 \sum_{m<n} \frac{1}{r_{mn}^{2}}  .  
\end{equation}

In $H_{4}$ we have terms of the order $1/r$ and $1/r^{3/2}$ but it turns out that they contribute only in 
higher  orders than $1/r^{2}$. The remaining terms give the answer 
\begin{eqnarray} \label{H4c}
\lefteqn {\langle\cdot|H_{4}|\cdot\rangle  \simeq \frac{7}{2}\sum_{p\ne m} \sum_{m \ne n} \frac{1}{r_{pm}r_{mn}} 
+} \nonumber \\
&&+ \frac{1}{2} \sum_{p\ne m \ne n \ne p} \left[  
\frac{2(\lambda_{m}-\lambda_{n})(\lambda_{p}-\lambda_{m})}{r_{mn}r_{pm}(\lambda_{p}-\lambda_{n})^{2}} +
 \frac{r_{pm}^{2} + r_{mn}^{2}}{r_{mn}r_{pm}(\lambda_{p}-\lambda_{n})^{2}} \right] . 
 \end{eqnarray} 

The correction term that gives a contribution of the order $1/r^{2}$ is 
\begin{eqnarray} \label{corc}
\lefteqn{-\langle\cdot |\left( -\frac{i}{2}f_{IAB}\phi_{I}^{a}\Lambda_{A}\Gamma^{a}\Lambda_{B} \right) 
\frac{1}{H_{osc}}\left( -\frac{i}{2}f_{JCD}\phi_{J}^{b}\Lambda_{C}\Gamma^{b}\Lambda_{D} \right) |\cdot\rangle = } 
\nonumber \\
&=&  8\sum_{p\ne m \ne n \ne p} \left[ - \frac{1}{r_{mp}(r_{pm}+r_{mn}+r_{np})} + 
\frac{(\lambda_{n}-\lambda_{m})(\lambda_{p}-\lambda_{n})}{r_{mp}r_{pn}r_{nm}(r_{mp}+r_{pn} + r_{nm})} 
\right] - \nonumber \\
&-& 16\sum_{m<n} \frac{1}{r^{2}_{mn}} . 
\end{eqnarray} 
Evidently the last term in (\ref{corc}) cancels with the analogous terms from (\ref{H0c}) and (\ref{H4c}). 
Those terms are similar to the ones arising in $SU(2)$ computation (see 
\cite{PfW}). The remaining terms  look, at the first glance, as if they can hardly cancel each other.   To see that 
this indeed happens we rewrite them in terms of $\lambda_{n}$ variables only, using the fact that 
$$r_{mn} = \lambda_{m} - \lambda_{n} + {\cal O}\left(\frac{1}{r}\right) \, , m<n  . $$  
Then (\ref{H4c}) contributes 
$$ 
 \sum_{m<p<n} \left( \frac{7}{(\lambda_{m}-\lambda_{p})(\lambda_{m}-\lambda_{n})} 
+ \frac{9}{(\lambda_{m}-\lambda_{p})(\lambda_{p}-\lambda_{n})}     + 
\frac{7}{(\lambda_{p}-\lambda_{n})(\lambda_{m}-\lambda_{n})} \right) . 
$$ 
The same procedure carried on the first two terms in (\ref{corc}) yields 
$$ 
-16\sum_{m<p<n} \left(  \frac{1}{(\lambda_{m}-\lambda_{p})(\lambda_{m}-\lambda_{n})} 
    + \frac{1}{(\lambda_{p}-\lambda_{n})(\lambda_{m}-\lambda_{n})} \right) .
$$ 
One readily checks that the sum of these two contributions vanishes. 
Hence, the outcome of our computation is given by the formula 
$$ 
H_{eff} = -\frac{1}{2}\frac{\partial^{2}}{\partial D_{i}^{\mu}\partial D_{i}^{\mu} } \, .
$$ 

In  appendix B we collected some formulae which we found useful for the computation described above.   


\section{Discussion }
First we would like to explain how the present results can be compared with 
those obtained in \cite{HS} for the $SU(2)$ gauge group.
 The authors of \cite{HS} use a gauge invariant variable $R$ to specify 
the asymptotic region. Namely,  $R$ is one of the eigenvalues of matrix 
$$ 
\Phi_{ab} = \phi^{\mu}_{a}\phi^{\mu}_{b} . 
$$ 
Here lower $a$ and $b$ are $SU(2)$ indices that run from $1$ two $3$.
Without loss of generality we can assume that index $3$ corresponds to the Cartan subalgebra and 
the basis $\phi_{a}^{\mu}$ is such that the matrix $\Phi$ is diagonal (any element of $SU(2)$ can be taken as 
the Cartan generator). Then, clearly,  
\begin{displaymath}
\Phi= \left( \begin{array}{ccc} 
*&0&0 \\
0&*&0 \\
0&0& r^{2} \end{array} \right)
\end{displaymath}
where $r^{2} = \phi_{3}^{\mu}\phi_{3}^{\mu} $ coincides with $r_{12}^{2}$ in our old notations. 
Hence, in this basis $R$ coincides with $r$. 

The result of \cite{HS} for the effective Hamiltonian (of radial degrees of freedom) is 
\begin{equation} \label{hHS} 
H_{eff} = -\frac{1}{2}\frac{d^{2}}{dR^{2}} - \frac{5}{R}\frac{d}{R} - \frac{4}{R^{2}} 
\end{equation}
while the measure on  Hilbert space is $R^{10}dR$. 
Composing this operator with $R$ from the left and with $R^{-1}$ from the right 
we get the radial part of the Laplacian in 
9 dimensions with the radial measure $R^{8}dR$. Therefore, the radial parts of our 
effective Hamiltonian and 
the one found by Halpern and Schwartz are the same. It is not hard to check that the angular 
dependence is the  same as well.

Once we have the effective Hamiltonian, we can find the asymptotic expression  for 
$P|\Psi \rangle = |\cdot \rangle |\Psi(D_{i} , \Lambda_{i})\rangle $ and then,
using (\ref{QPsi}) and (\ref{pert1}), one can get the  asymptotic form of the whole 
state vector $|\Psi\rangle $. Asymptotic solutions to $H_{eff} |\Psi(D_{i} , \Lambda_{i})\rangle = 0$ 
have a basis of the form 
$$
 D_{1}^{-7-l_{1}}Y_{l_{1}}(D_{1}^{\mu}) \cdot \ldots \cdot  
D_{N-1}^{-7-l_{N-1}}Y_{l_{N-1}}(D_{N-1}^{\mu}) |\Lambda_{1} , \ldots , \Lambda_{N-1} \rangle 
$$ 
where $Y_{l_{i}}$ are SO(9) spherical harmonics. 
Further constraints of the supersymmetry, SO(9)-invariance and  invariance with respect 
to the Weyl group of $SU(N)$ (i.e. the permutation group $S_{N}$) should be put on 
$|\Psi(D_{i} , \Lambda_{i})\rangle$ to single out the  candidates for the ground state.

Another comment we would like to make here is about the computation of Witten index. 
In paper \cite{SS} it was shown that Witten index for the model at hand has two contributions none of 
which can be made vanishing by the choice of the limiting procedure. The following expression for the 
correction (or surface) term was found 
$$ 
{\cal I}_{surf} = -\frac{1}{2} \int_{N_{F}(R)} tr e_{n} (-1)^{F} QW'  . 
$$
The main ingredient in this formula is $W'$ - the approximate Green's function in the limit of 
large separation. The authors of \cite{SS} showed that one can take a free propagator for $W'$ 
in the $SU(2)$ theory and gave the complete calculation of index in this case. In \cite{GG1} the 
surface term was calculated under a similar assumption for $W'$ in $SU(N)$ theory. 
 The present paper justifies  this assumption and therefore completes the proof that the full 
Witten index for the $SU(N)$ case is equal to 1.

\begin{center} {\bf Acknowledgements} \end{center}

I would like to thank M.~Halpern and C.~Schwartz for helpful and stimulating discussions. 
I also  want to express my gratitude to A.~Schwarz for 
his  constant support and  interest in my work.


\appendix
\section{} 
In this appendix we give expressions for  the  summands appearing in  $H_{4}$ in terms of 
 $D_{i}^{\mu}$, $\Lambda_{i\alpha}$, $\phi^{a}_{mn}$, $\Lambda_{\alpha}^{mn}$
$$ 
\frac{1}{4}f_{AIJ}f_{AKL}\phi_{I}^{a}\phi_{J}^{b}\phi_{K}^{a}\phi_{L}^{b} = 
\sum_{m\ne n \ne p \ne q \ne m } ( \phi_{mn}^{a}\phi_{np}^{a}\phi_{pq}^{b}\phi_{qm}^{b} - 
\phi_{mn}^{a}\phi_{np}^{b}\phi_{pq}^{a}\phi_{qm}^{b}) \eqno (A.1)
$$
$$ 
  \frac{1}{2}f_{AiJ}f_{AKL}(D^{a}_{i}\phi_{J}^{b}-D_{i}^{b}\phi_{J}^{a}) \phi_{K}^{a}\phi_{L}^{b} 
= $$
$$ 
=\sum_{m\ne n\ne p\ne m} \frac{1}{\sqrt{2}} ((\lambda_{m}^{a}-\lambda_{n}^{a})\phi_{mn}^{b}-
(\lambda_{m}^{b}-\lambda_{n}^{b})\phi_{mn}^{a})\phi_{np}^{a}\phi_{pm}^{b} 
\eqno (A.2) 
$$
$$
 -\frac{1}{2}f_{AiJ}f_{AkL}D_{i}^{a}D_{k}^{b}\phi_{J}^{b}\phi_{L}^{a} =
\sum_{m \ne m} \frac{1}{2}
(\lambda_{n}^{a}-\lambda_{m}^{a})(\lambda_{m}^{b}-\lambda_{n}^{b})\phi_{mn}^{b}\phi_{nm}^{a} 
\eqno (A.3)
$$
$$
 - \frac{i}{2}f_{IAB}\phi_{I}^{a}\Lambda_{A}\Gamma^{a}\Lambda_{B} =
-i\sum_{m\ne n; p \ne q} [E_{mn} , E_{pq}]_{i} \phi_{mn}^{a} \Lambda^{pq}\Gamma^{a}\Lambda_{i}
+
$$
$$  +\frac{1}{\sqrt{2}} \phi_{mn}^{a}\Lambda^{np}\Gamma^{a}\Lambda^{pm}  
\eqno (A.4)
$$
$$
\frac{1}{2}(w^{t}w)_{IJ}\hat L_{I}\hat L_{J} = \sum_{m\ne n} \frac{-1}{2(\lambda_{n}-\lambda_{m})^{2}} 
\hat L_{nm}\hat L_{mn} \eqno (A.5)
$$
where 
$$
\hat L_{nm} = \frac{i}{\sqrt{2}} \sum_{p}(\phi_{mp}^{a} \frac{\partial}{\partial \phi_{np}^{a}} - 
\phi_{pn}^{a} \frac{\partial}{\partial \phi_{pm}^{a}} + 
$$
$$
+\frac{i}{\sqrt{2}} \Lambda_{\alpha}^{mp}\Lambda_{\alpha}^{pn} ) - \sum_{p\ne q}[E_{nm}, E_{pq}]_{j} 
(D_{j}^{a}\frac{\partial}{\partial \phi_{qp}^{a}} - \phi^{a}_{pq}\frac{\partial}{\partial D_{j}^{a}} 
+ \Lambda_{j}\Lambda^{pq} )  
\eqno (A.6) 
$$

\section{}
When working with bosonic oscillators it is convenient to introduce the creation and annihilation 
operators as  follows 
$$ 
 \phi_{mn}^{b} = \frac{ a_{mn}^{b} + (a^{b}_{nm})^{\dagger} }{\sqrt{2r_{mn}}} 
$$ 
$$ 
\frac{\partial}{\partial \phi_{mn}^{b}} = \frac{ a_{nm}^{b} - (a_{mn}^{b})^{\dagger}  }{\sqrt{2/r_{mn}}} 
$$ 
$$
[a^{b}_{mn} , (a^{c}_{pq})^{\dagger} ] = \delta^{ab}\delta_{mp}\delta_{nq} \enspace m\ne n , p\ne q  . 
\eqno (B.1)
$$
Since both bosonic and fermionic oscillators ground states depend on eigenvalues $\lambda_{m}^{\mu}$,
a certain care needs to be taken when calculating the averages of differential operators over the oscillators. 
The following formulae are useful 
$$ 
\frac{\partial a^{b}_{nm}}{\partial \lambda_{k}^{\mu} } = 
(\delta^{mk} - \delta^{nk}) \frac{(\lambda_{m}^{\mu} - \lambda_{n}^{\mu})}{2r_{mn}^{2}} (a^{b}_{mn})^{\dagger} 
\eqno (B.2)
$$ 
$$
\frac{\partial (a^{b}_{nm})^{\dagger}}{\partial \lambda_{k}^{\mu} } = 
(\delta^{mk} - \delta^{nk}) \frac{(\lambda_{m}^{\mu} - \lambda_{n}^{\mu})}{2r_{mn}^{2}} a^{b}_{mn} 
\eqno (B.3)
$$   
 
$$ 
    [ \frac{\partial}{\partial \lambda_{k}^{\mu}} , |\cdot\rangle ] = 
-\sum_{m<n} (\delta_{mk} - \delta_{nk} ) [ \frac{(\lambda_{m}^{\mu} - \lambda_{n}^{\mu} )}{2r_{mn}^{2}} 
(a^{b}_{nm})^{\dagger}(a^{b}_{mn})^{\dagger} +  
$$
$$
+ \frac{1}{2}\left( \delta^{\mu 9} + 
\frac{(1-\delta^{\mu 9} ) (\lambda_{m}^{\mu} - \lambda_{n}^{\mu} ) }{r_{mn} + \lambda_{m} - \lambda_{n}} 
\right) \frac{(\lambda_{m}^{a} - \lambda_{n}^{a})}{r_{mn}^{2}} (\tilde \Lambda^{mn})^{\dagger} \Gamma^{a} 
\tilde \Lambda^{mn} -   
$$
$$ 
- \frac{1-\delta^{\mu 9}}{2r_{mn}}(\tilde \Lambda^{mn})^{\dagger} \Gamma^{\mu}  
\tilde \Lambda^{mn} ] | \cdot \rangle  . \eqno  (B.4)
 $$

Fermionic quadratic correlator is 
$$ 
\langle 0_{F} | \Lambda_{\alpha}^{mn}\Lambda_{\beta}^{pq} | 0_{F}\rangle = \frac{1}{2}\delta^{mp}\delta^{nq} 
\left( 1 - \gamma_{9} \frac{\lambda_{m} - \lambda_{n}}{r_{mn}} \right)_{\alpha\beta} 
\eqno (B.5)
$$ 
for any $m \ne n $.

Note that $spin(9)$-rotated fermions $\tilde \Lambda^{mn}$ depend on $\lambda_{m}^{\mu}$. 
The following correlator thus appears in computations 
$$ 
\langle 0_{F} | \frac{\partial}{\partial \lambda_{k}^{\mu} } (\tilde \Lambda^{mn})^{\dagger} \Gamma^{a} 
\tilde \Lambda^{mn} | 0_{F} \rangle = -\frac{1}{2} Tr ( \Gamma^{9} 
[ R^{mn}\frac{ \partial (R^{mn})^{t} }{\partial \lambda_{k}^{\mu}} , \Gamma^{a} ] ) . 
\eqno (B.6)
$$


\end{document}